\documentclass[11pt,a4paper]{article}
\usepackage[hyperref]{acl2021}
\usepackage{times}
\usepackage{latexsym}

\usepackage{tikz}
\usetikzlibrary{shapes.misc,shadows}
\newcommand{\yellowcl}[1]{%
  \tikz[baseline=(char.base)]\node[anchor=south west, draw, fill=yellow, rectangle, rounded corners](char){#1} ;}

\usepackage{microtype}
\usepackage{todonotes}
\usepackage{array}
\newcolumntype{L}[1]{>{\raggedright\let\newline\\\arraybackslash\hspace{0pt}}m{#1}}
\newcolumntype{C}[1]{>{\centering\let\newline\\\arraybackslash\hspace{0pt}}m{#1}}
\newcolumntype{R}[1]{>{\raggedleft\let\newline\\\arraybackslash\hspace{0pt}}m{#1}}
\newcommand\Mark[1]{\textsuperscript#1}
\usepackage{enumitem}
\usepackage{makecell}
\usepackage{amsmath}
\usepackage{tabularx}
\usepackage{soul}

\aclfinalcopy 
\def\aclpaperid{226} 

\title{{P}hoto{C}hat: A Human-Human Dialogue Dataset with Photo Sharing Behavior for Joint Image-Text Modeling}

\author{Xiaoxue Zang\Mark{1}, Lijuan Liu\Mark{1}, Maria Wang\Mark{1}, Yang Song\Mark{2}\Thanks{ Research conducted while working at Google.}, 
{\bf Hao Zhang\Mark{1}}, {\bf Jindong Chen\Mark{1}} \\
\Mark{1} Google Research, \Mark{2} Kuaishou Technology \\
\Mark{1}\{xiaoxuez, 
lijuanliu, mariawang, haozhangthu, jdchen\}@google.com, \\ 
\Mark{2} yangsong@kuaishou.com}

\date{}

\begin{document}
\maketitle
\begin{abstract}
We present a new human-human dialogue dataset - PhotoChat, the first dataset that casts light on the photo sharing behavior in online messaging. PhotoChat contains 12k dialogues, each of which is paired with a user photo that is shared during the conversation. Based on this dataset, we propose two tasks to facilitate research on image-text modeling: a photo-sharing intent prediction task that predicts whether one intends to share a photo in the next conversation turn, and a photo retrieval task that retrieves the most relevant photo according to the dialogue context. In addition, for both tasks, we provide baseline models using the state-of-the-art models and report their benchmark performances. The best image retrieval model achieves 10.4\% recall@1 (out of 1000 candidates) and the best photo intent prediction model achieves 58.1\% F1 score, indicating that the dataset presents interesting yet challenging real-world problems. We are releasing PhotoChat to facilitate future research work among the community. 

\end{abstract}

\section{Introduction}\label{sec:intro}


As instant messaging tools gain enormous popularity in the recent decades, sharing photos as an approach to enhance the engagement of an online messaging conversation has become a pervasive routine communicative act~\cite{Lobinger2016}. A survey conducted in 2010 reveals that 74\% of teenagers in the US reported messaging a photo or video using their cell phone~\cite{Lenhart2010TeensAM}. In Britain, almost 70\% of the internet users shared photos in 2013~\cite{Dutton2013}. Considering the proliferation of photo sharing, it's desirable to have an intelligent system that can assist users efficiently engaging in this process, i.e. suggesting the most relevant photos in correct timings. In order to achieve this goal, the intelligent system is expected to not only understand how humans communicate with each other, e.g. the natural language human speak, but also perceive images as human do. How to facilitate building such multimodal system is the goal of this paper.


\begin{figure}[t]
  \centering
  \setlength{\abovecaptionskip}{-10pt}
  \includegraphics[width=\linewidth]{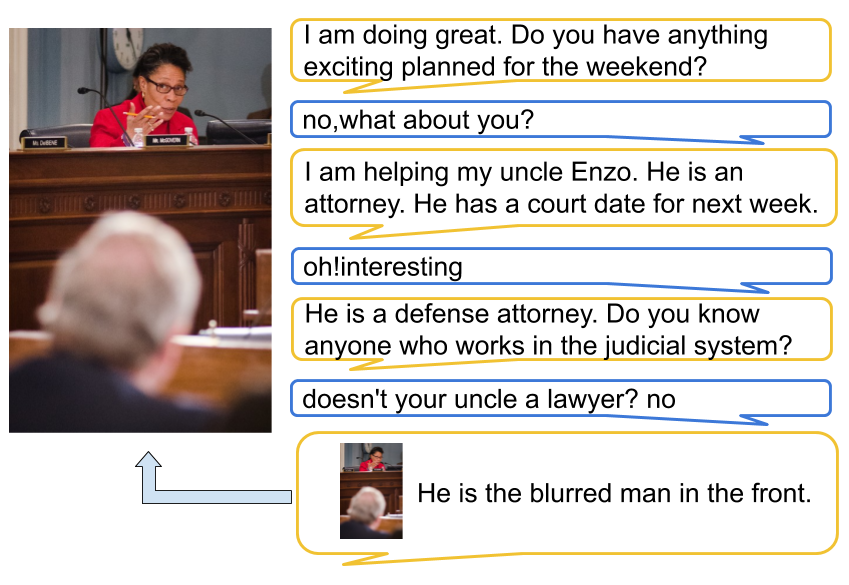}
    \setlength{\belowcaptionskip}{-20pt}
  \caption{An example of how people share photos in a daily conversation.}
  \label{fig:front}
\end{figure}

Though recently many image-text tasks have been proposed and are being actively studied to bridge language and vision, the majority of them are formulated as choosing or composing the text based on the understanding of given images, e.g. image captioning \cite{anderson2018bottomup}, visual question answering \cite{vqa}, visual commonsense reasoning \cite{zellers2019vcr}, and image-grounded dialogue generation \cite{shuster-etal-2020-image}. Contrary to these tasks, the photo sharing task focuses on the reverse process, i.e. selecting the image based on the understanding of text, as well as proposing different and unique challenges. 

Firstly, different from the above popular multimodal tasks, in photo-sharing task, the dialogue doesn't often explicitly mention the main visible content in the image. Instead of the main object of the photo, sometimes the background story, complemented by human imaginations, can be the focus of the chat. Figure~\ref{fig:front} shows such an example, in which the person who shares the photo describes the event location ``court" and the occupation ``attorney" instead of the main object ``lady" in the image. Secondly, the dialogue is not guaranteed to be relevant to the image. For instance, it often contains greetings and chit-chats of other topics, as the first two turns in Figure~\ref{fig:front} shows. In order to suggest the relevant photo, a smart system needs to decide which part of the dialogue can be used for suggesting the image. In contrast, in the traditional image-text tasks, the correct text is designed to be highly correlated with the image and has few distracting content. These photo sharing characteristics makes inferring the connection between the image and textual utterances challenging. 


To highlight these challenges, we create PhotoChat - a human-human dialogue dataset in which one photo is shared from one person to the other during the conversation\footnote{ https://github.com/google-research/google-research/tree/master/multimodalchat/}. It is, as far as we know, the first dataset that captures the photo sharing activities. 
We selected images from OpenImage V4 dataset~\cite{OpenImages} as shared photos and used crowdsourcing plugins to generate 12,286 dialogues with an average of 10 turns per dialogue. During the dialogue collection, the photo is only visible to the side who is instructed to share the photo and then to both sides after it is being shared. Based on the collected dataset, we propose two tasks that are essential for building a photo suggest system: photo-sharing intent prediction task that predicts whether one intends to share the photo in the next conversation turn, and dialogue-based image retrieval task that retrieves the most relevant photo given the dialogue context. For both, we build baseline models, report and analyze their performances. The best photo-sharing intent prediction baseline model achieves 58.1\% F1 score with 58.2\% precision and 57.9\% recall. The best cross-attention image retrieval model achieves 10.4\% recall@1 out of 1000 candidates. We also propose a dual-encoder model that leverages object labels to encode image features, which achieves the best performance among all the models w/o cross-attention mechanisms.

In summary, our main contributions are:
\begin{itemize}[align=left,labelsep=0em,noitemsep,topsep=0.2pt]
    \item We create the first human-human dialogue with photo sharing acts via crowd-sourcing. 
    \item We propose two new tasks to promote building an intelligent photo suggest system.
    
    \item We build baseline models and provide benchmarks for the new tasks. Our proposed image retrieval model outperforms all the prior models w/o cross-attention mechanisms. We implement comprehensive analysis and ablation study to provide more insights.
\end{itemize}

\section{Related Work}
With the recent advances in deep learning, plenty of image-text datasets have been created and new image-text tasks are proposed based on them. These datasets have greatly stimulated the development of joint image-text models. In this section, we review the widely used image-text datasets and the state-of-the-art (SOTA) approaches for solving the image-text problems.

\subsection{Image-text Dataset}
Image-captioning datasets are first widely used for joint image-text modeling. 
MSCOCO~\cite{miscoco_lin_2014} and Flickr30k \cite{young-etal-2014-image} that both contain five written caption descriptions for each image are the representative ones used for automated caption generation and cross-modal retrieval tasks. Conceptual Caption~\cite{sharma2018conceptual} is yet another popular image caption dataset but contains an order of magnitude more images than MSCOCO. Because image captions usually only describe the main objects in the image and omit details, to facilitate understanding details of an image along with the reasoning behind them, \citet{vqa} introduced VQA which contains three question answer pairs for each image. A further work is VCR \citep{zellers2019vcr} that not only requires a model to answer the question derived from the image but also provides a rationale explaining why its answer is right. It was created to teach the model to learn higher-order cognition and commonsense reasoning about the world.

Compared to the work above, Image-Chat \cite{shuster-etal-2020-image} and IGA \cite{mostafazadeh2017imagegrounded}, which focus on the dialogues grounded in the image, are the most related work to ours. IGA includes 4k dialogues where each contains an image with a textual description of it, along with the questions and responses around the image. Due to its small scale, IGA can only be used for evaluation. Image-Chat is a larger scale dataset that consists of 202k image-grounded dialogues. However, both of them were created by asking the crowd workers to talk about a shared image to generate engaging conversation, which is different from the scenario of photo sharing where only one side can access the photo at the start of the conversation. Thus, neither can be used to build a photo-suggest system. In our work, we build a new dataset that highlights the challenges of building a photo-suggest system and is the first of its kind to the best of our knowledge.
\subsection{Image-text Modeling}
As the challenge for the photo-suggest system is to retrieve the most relevant image based on the textual utterances, we only review the related work on cross-modal retrieval.

Many models have been proposed for image-caption retrieval where one is required to retrieve the most relevant caption given an image or vice versa. The typical architecture consists of two separate encoders for image and text to first generate visual and textual embeddings. On top of them, a fusion layer, which can simply be a dot product, is used to generate the relevance score for each pair \cite{NIPS2013_5204, kiros2014unifying, parekh2020crisscrossed, karpathy2015deep, faghri2018vse++}. Then a triplet ranking loss or cross-entropy loss is employed to learn the latent visual-semantic alignment. VSE++ \cite{faghri2018vse++} emphasizes on the hardest negatives by using the max of the hinge loss as the objectives and yielded a significant performance improvement. Stacked Cross Attention Network (SCAN) \cite{lee2018stacked} further improves the performance by introducing the cross attention between image regions and word features. Recently, cross-modal transformer based architecture that are pretrained on large-scale image-text datasets via self-supervised learning has shown great advantages in bridging visual and textual embeddings. Multiple concurrent work \cite{lu2019vilbert, chen2020uniter, li2019unicodervl} have refreshed the best records on the benchmark datasets for the image-text retrieval tasks.

\section{Dataset Creation}

\begin{figure*}[t]
\centering 
   \small
\begin{minipage}{16cm}\vspace{0mm}    \centering
   \begin{tabularx}{\linewidth}{c|c|c}
   \hline \hline
   \textbf{Good Example} & \textbf{Good Example} & \textbf{Bad Example} \\ \hline
   \raisebox{-\height}[0pt][80pt]{\includegraphics[width=0.2\textwidth]{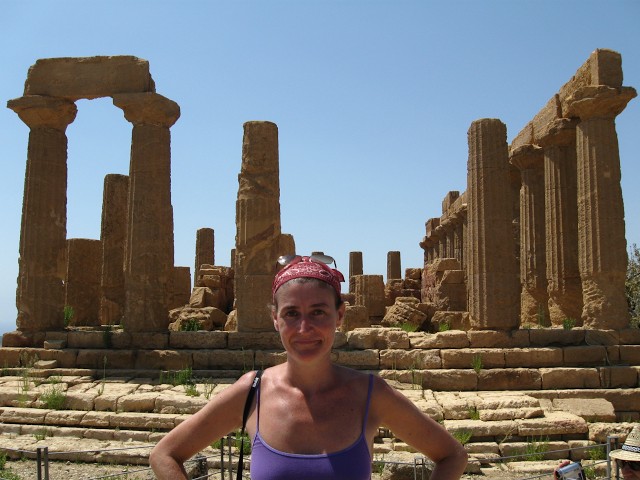}} & \raisebox{-\height}[0pt][80pt]{\includegraphics[width=0.15\textwidth]{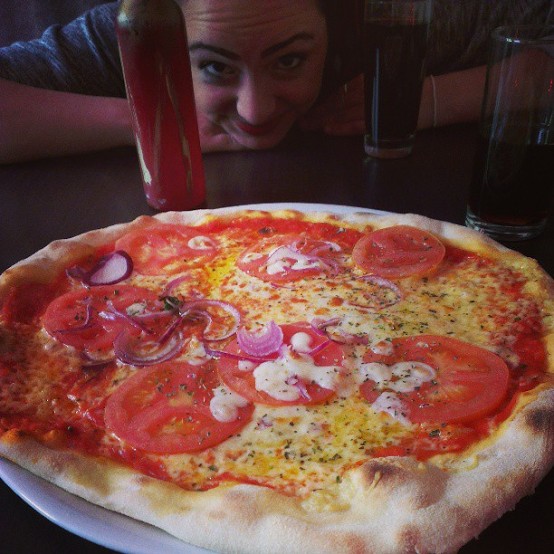}} &  \raisebox{-\height}[0pt][80pt]{\includegraphics[width=0.2\textwidth]{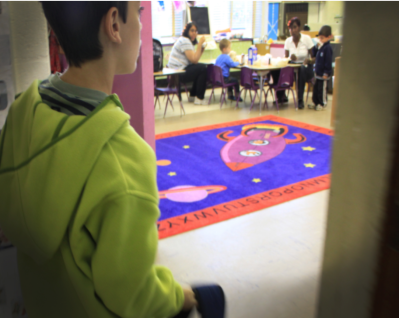}} \\
    \makecell[Xt]{
    \textbf{A:} hows it going? \\
    \textbf{B:} just got back from vacation!! \\
    \textbf{A:} How was vacation? did you have fun? \\
    \textbf{B:} It was exciting! I took my granddaughter to Greece and we saw so many beautiful ruins! \\
    \textbf{A:} oh wow! Greece, that's amazing. I bet you got amazing pictures of the ruins \\
    \textbf{B:} Yeah, we saw ancient temples and battlefields \\
    \textbf{B:} \yellowcl{Share the photo} \\
    \textbf{A:} Wow! that's a great photo. you should post it on Insta too.\\
    \textbf{B:} Great idea! Thanks!} & 
    \makecell[Xt]{
    \textbf{A:} hey guess what i'm doing now ?? \\
    \textbf{B:} What are you up to today? \\
    \textbf{A:} i'm preparing a pizza for the first time i include tomatoes,onions and so on \\
    \textbf{B:} Wow, you must be daring! Whoever taught you should have been confident on your progress. \\
    \textbf{A:} hey..... i'm almost done \\
    \textbf{B:} Must be yummy> \\
    \textbf{A:} wanna see my preparation? \\
    \textbf{A:} \yellowcl{Share the photo}
    }& \makecell[Xt]{
    \textbf{A:} How are you? \\
    \textbf{B:} I'm doing well. I've been watching Netflix because I can't go outside. \\
    \textbf{A:} Yeah, same here. Which show? \\
    \textbf{A:} And actually, I just found this picture of someone who should be a photographer. \\
    \textbf{B:} The office has been my go to. \\
    \textbf{B:} Really? Share the photo to me. \\
    \textbf{A:} \yellowcl{Share the photo} \\
    \textbf{B:} Whoa! You were totally right \\
    \textbf{A:} It's a boy in neon green who I think wants to take photos in academic settings. \\
    \textbf{B:} This photo is so cool
    }\\ \hline \hline
    \end{tabularx}
\vspace{-2mm}
\caption{Examples of PhotoChat dataset. The first two examples are included in the dataset while the last example is excluded in the verification step. \yellowcl{Share the photo} denotes the photo sharing act.}\label{fig:data_ex}
\end{minipage}
\vspace{-3mm}
\end{figure*}

We select photos from Open Image Dataset V4 (OID)~\cite{OpenImages} and collect open-ended conversations on Amazon Mechanical Turk. Below describes the detailed image filtering, conversation generation, and data verification steps to ensure data quality.
\subsection{Image-based Filtering}
Since OID is large-scale and comprehensive, it contains images that are unlikely to be shared in the daily dialogue, such as images only about remote controls or fire hydrants. To create a dataset that is close to the reality, we filter images based on the annotated object labels provided with OID.

Based on our investigation of the image-grounded dialogues and daily experiences, photos about four themes are commonly shared: people, food, animal, and product (in the shopping scenario), which are our focus in the dataset creation. From all the 600 object labels that appear in OID, we first enlist the labels that both belong to one of the four themes and have a high chance to appear in the commonly-shared photos. Labels like ``traffic light", ``nail", and ``reptile" are excluded and labels like ``girl", ``bagel", and ``camera" are included. This process selects 89 object labels (Appendix). We then generate an image pool by selecting those that contain any of the objects in the list. Note that for the objects of the people category, we add another criteria that it must be the main object, i.e. neither positioned in the margin of the image\footnote{Center of the object is located within 0.1 of the image width/height to the border.} nor extremely small \footnote{Object width/length $<$ 0.3 $\times$ (image width/length).} to exclude images that only have people as the background. Images are randomly selected from the image pool to generate conversations in the next step.

\subsection{Conversation Generation}
We randomly assigned two crowd workers to generate a conversation based on a given image. The image comes with an image description which presents the list of objects labels in the image. When the image contains humans, we assign a random name and relationship to one of the humans to help the workers refer to it and unfold the story. They are instructed to imagine talking with their friend. At the start of the task, only one side has access to the image and is instructed to drive the dialogue until it is fit to share the image with the other (website interfaces are shown in the Appendix). It is not restricted that they must message alternatively but the worker with the photo can't share the photo until the total number of the conversation turns reaches five. After sharing the photo, they can continue to chat until they wish to end the conversation and submit the dialogue. 

\subsection{Image\&text-based Verification}
Lastly, we use another set of in-house professional crowd workers to filter out the invalid dialogues generated in the above step. Dialogues are discarded if the association between the image and the dialogue is in-evident before the photo sharing act or the content is unnatural, contains inappropriate words, too many typos or broken English.
Figure~\ref{fig:data_ex} displays examples of qualified and unqualified data. Note that the third unqualified dialogue can happen in a real conversation, yet the content/event of the image is not mentioned until the photo being shared, making it impossible for a model to learn the connection between the dialogue and the images and to suggest a photo in advance. Such dialogues are removed from the dataset in this step.

\section{Dataset Statistics}

\begin{table*}[t]
    \caption{PhotoChat statistics. Table shows the aggregated numbers. From left to right starting from the second column, the name of each column means ``the unique number of images", ``the number of dialogues", ``the number of dialogues about people/food/animal/product", ``the number of turns", ``the number of turns when counting consecutive turns of the same speaker as one turn", and ``the number of tokens". Turns in which photos are shared are excluded in the calculation.}
    \label{tab:split_stats}
    \vspace{-2mm}
    \centering
    \begin{tabular}{C{0.8cm}|C{1.15cm}|c|C{1.15cm}|C{1.15cm}|C{1.3cm}|C{1.3cm}|c|c|c} \hline
        \textbf{split} & \textbf{unique img \#} & \textbf{dial \#} & \textbf{people dial \#} & \textbf{food dial \#} & \textbf{animal dial \#} &  \textbf{product dial \#} & \textbf{turn \#}  & \textbf{turn* \#}  & \textbf{token \#} \\ \hline
        train & 8,917 & 10,286 & 6,376 & 4,465 & 1,072 & 884 & 130,546 & 97,586 & 827,154 \\ \hline
        dev & 1,000 & 1,000 & 606 & 424 & 87 & 109 & 12,701 & 9,533 & 80,214 \\ \hline
        test & 1,000 & 1,000 & 615 & 419 & 90 & 108 & 12,852 & 9,590 & 80,847 \\ \hline \hline
        total & 10,917 & 12,286 & 7,597 & 5,308 & 1,249 & 1,101 & 156,099 & 116,709 & 988,215 \\ \hline \hline
    \end{tabular}
    \vspace{-5mm}
\end{table*}

The collected dataset consists of 10,917 unique images and 12,286 dialogues. One image is shared in each dialogue. Based on the object labels of the shared image, we classify the dialogues into four categories: people, food, animals, and daily products. We split the dialogues into 10,086 train, 1,000 dev, and 1,000 test sets while keeping roughly the same distribution of the category across the splits. The detailed statistics of each split and in total are shown in Table~\ref{tab:split_stats}. Note that the dialogue can have multiple category labels. For instance, if the shared image is about a girl playing with dogs, the dialogue belongs to both people and animals categories. Thus, the sum of the dialogues of each category (people/animal/food/product dial \#) exceeds the total number of the dialogues (dial \#) in the table. In addition, some images in the training set are used in multiple dialogues.

Based on the statistics in the table, the average number of turns per dialogue is 12.7 and the average number of tokens per turn is 6.3. Since two sides are not restricted to speak alternatively, if the consecutive turns from the same side are combined as one turn, which is the conventional setting of other dialogue datasets, the average number of turns per dialogue and the average number of tokens per turn become 9.5 and 8.5. On average, people converse for 7 turns before sharing the photo. 

\section{Task Definition}
We decompose the problem of building a smart photo-suggest system into two separate tasks. The first is to detect if the user has the intent to share the photo in the next turn, which we call photo-sharing intent prediction task. The second is to retrieve the photo based on the dialogue context, which we call image retrieval task. 
Below describes the formal formulation of the problem settings.

Let $P = \{p_1, p_2, ..., p_M\}$ be the photo set where each $p_i = (a_i, l_i)$, $i\in[1, M]$ consists of image $a_i$ and a list of objects $l_i$ in it. Given the dialogue $D = \{t_1, ..., t_h, p_k, t_{h+1}, ..., t_N\}$ where two participants speak alternatively, $t_j$ ($j\in[1,N]$) and $p_k\in P$ respectively represent the utterance of turn $j$ and the shared image. $t_h$ is the turn immediately before a photo sharing act. We also define the speaker information $S = \{s_1, s_2, ..., s_N\}$ where $s_j$ $(j\in[1, N])$, either $0$ or $1$, denotes the speaker of turn $j$. 

\textbf{Photo-sharing intent prediction:} The goal of the intent prediction task is to predict whether a photo will be shared in the next turn for any $t_j$ given all the turns before. In equation, it's formulated as a binary classification task:
\begin{align}\label{eq:class}
    \small
    \forall j \in [1, h], C(t_{1:j}, s_{1:j}) \in \{0, 1\}, 
\end{align}
where $C$ is the intent prediction model taking the utterances and the speaker information of all the previous turns as the input and outputs a binary value. In the above case, it should only predicts $1$ when $j = h$, otherwise 0. Note that whether the model make use of all the previous turns and the speaker information depends on the model design. We use F1 score, precision, and recall as the evaluation metrics for this task.

\textbf{Image retrieval:} Under the same settings, model $R$ of the image retrieval task is expected to correctly retrieve $p_k$ from $P$ given the dialogue:
\begin{align}
    R(t_{1:h}, s_{1:h}, P) \in [1, M].
\end{align}
During training, the candidate pool $P$ is usually comprised of in-batch images while 
during evaluation, $P$ contains all images in the test set. Following \citet{karpathy2015deep}, we use Recall@K (R@K), computed as ``the fraction of times a correct item was found among the top K results" as the evaluation metrics. Specifically, we choose R@1, R@5, and R@10, as well as the sum of them which we denote as ``sum(R@1, 5, 10)" to evaluate the models.

\section{Baselines}
\subsection{Photo-sharing Intent Prediction Model}
To establish the baselines, we fine-tune three SOTA pretrained models - BERT~\citep{jacob2018bert}, ALBERT~\citep{lan2020albert}, and T5~\citep{2020t5}, as the pretrained models have achieved remarkable performance in many NLP tasks.

To adapt BERT and ALBERT to our settings, we concatenate all the previous turns ($t_{1:j}$ in Equation~\ref{eq:class}) by \texttt{[SEP]} and prepend the concatenated text with \texttt{[CLS]} to generate the input to the model. We use the speaker information $s_{1:j}$ as the segment id of the input. The output of \texttt{[CLS]} token is fed into two fully-connected layers, of which the output dimensions are respectively 128 and 2 to generate the final prediction. To utilize T5, we concatenate $t_{1:j}$ by \texttt{[SEP]} and prepend the text with ``predict share intent:" as the model input. We use cross entropy loss for all three models.


\subsection{Image Retrieval Model}
Our baselines consists of both statistical and neural network-based approaches, as elaborated below:
\begin{figure*}[t]
  \centering
  \includegraphics[width=\linewidth]{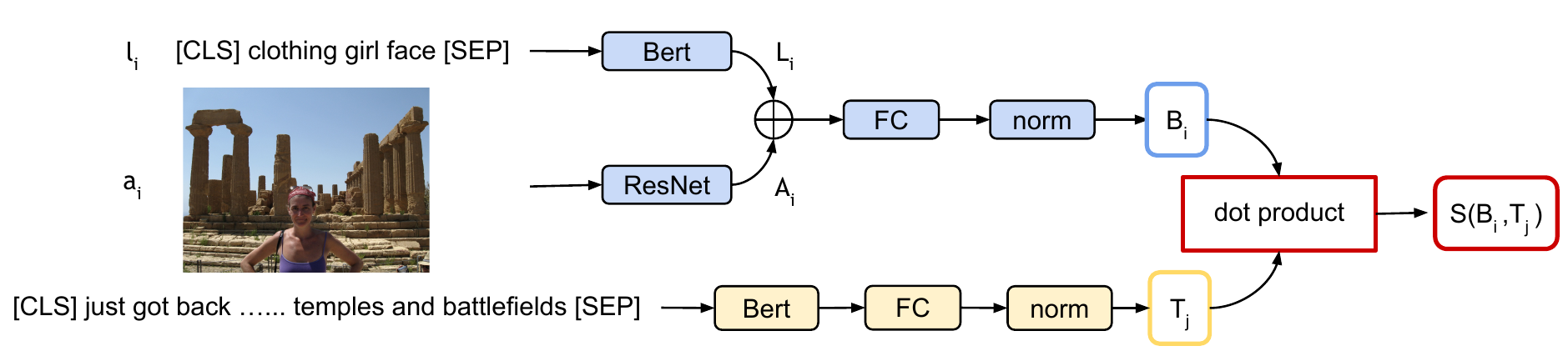}
    \setlength{\abovecaptionskip}{-10pt}
  \caption{Our dual encoder. The first dialogue in Figure~\ref{fig:data_ex} is used as the input example. Image and text are encoded separately to generate their embeddings. The dot product of them is then used to compute the similarity score.}
  \label{fig:model}
  \setlength{\belowcaptionskip}{-20pt}
  \vspace{-5mm}
\end{figure*}

\textbf{Dual encoder:}
We built a dual-encoder model similar to \citet{parekh2020crisscrossed, gillick2018endtoend}, which separately encodes image and text leveraging SOTA pre-trained models. Its entire architecture is shown in Figure~\ref{fig:model}.

To encode the image, for each $p_i = (a_i, l_i)$ we first resize the image $a_i$ to $224\times 224$ and feed it into a pretrained ResNet~\cite{he2016deep} to generate $A_i$. A pretrained BERT is used to encode $l_i$ to achieve the label embedding $L_i$ which is the output of \texttt{[CLS]} token. $L_i$ is concatenated with $A_i$ to generate the image embedding.
For encoding the dialogue context, we use a second pretrained BERT \cite{devlin2018bert}. Its input is the concatenation of all the prior utterances of the speaker who shares the photo. The output of \texttt{[CLS]} token is used as the contextual text embedding. 
 Two fully connected layers are then used to separately project image and text embeddings into a joint image-text embedding space of dimension $H$. Then, the dot product of the normalized image embedding $B_i$ and text embedding $T_j$ is used as the similarity score $S(B_i, T_j)$. Following \citet{young-etal-2014-image, gillick2018endtoend}, bidirectional in-batch sampled cross entropy loss is employed:
\begin{align*}
    l_{sm}(B_i, T_j) = - (S(B_i, T_j) - \log{\sum_{\hat{T_j}}{e^{S(B_i, \hat{T_j})}}}) \\ 
    - (S(B_i, T_j) - \log{\sum_{\hat{B_i}}{e^{S(\hat{B_i}, T_j)}}}),
\end{align*}
where $\hat{B_i}$ and $\hat{T_j}$ are the image embeddings and text embeddings of the other examples in the batch.

We also experiment with bidirectional in-batch hinge loss, defined as:
\begin{align*} 
l_{sh}(B_i, T_j) = \sum_{\hat{T_j}}[\alpha - S(B_i, T_j) + S(B_i, \hat{T_j})]_+\\ 
+ \sum_{\hat{B_i}}[\alpha - S(B_i, T_j) + S(\hat{B_i}, T_j)]_+,
\end{align*} 
where $\alpha$ is the margin parameter and $[x]_+ \equiv max(x, 0)$ . In our preliminary experiments, we observe
cross entropy loss works better and implement most experiments with cross entropy loss.

\textbf{VSE++:} VSE++~\cite{faghri2018vse++} is a simple and effective dual encoder model. It encodes the image and the text, which is the concatenation of all the previous utterances of the person who shares the photo in our case, separately by ResNet152~\cite{he2016resnet} and GRU~\cite{cho2014learning}. It is then followed by linear projections to map them into the joint embedding space. Finally, dot products of the normalized embeddings are used to compute the ranking scores. They innovatively make use of the hardest negatives, which are the negatives closest to the query, in the ranking loss function:
\begin{align*}
    l_{mh}(B_i, T_j) = [\alpha - S(B_i, T_j) + S(B_i, \hat{T_{j}^h})]_+ \\ 
+ [\alpha - S(B_i, T_j) + S(\hat{B_{i}^h}, T_j)]_+,
\end{align*}
where $\hat{T_{j}^{h}} = argmax(S(B_i, \hat{T_j}))$ and $\hat{B_i^{h}} = argmax(S(\hat{B_i}, T_j))$ are the hardest negatives.

\textbf{SCAN:} SCAN~\cite{lee2018stacked} is a full cross attention model that captures the fine-grained interplay between image regions and text tokens to infer image-text similarity. 
It uses fasterRCNN~\cite{ren2017fastrcnn} in conjucntion with ResNet-101 to compute image region embeddings and bidirectional GRU to achieve text embeddings. Same as VSE++, SCAN uses hard negatives in the triple ranking loss function. Though it beats VSE++ on the image captioninig tasks, it doesn't scale well to large-scale retrieval problems due to the high computational cost of cross attention.

\textbf{BM25:} BM25~\cite{Amati2009} is a probabilistic retrieval function widely used for document retrieval. To adapt it to our settings, we directly utilize the object labels of each image $l_j, j \in [1, m]$ as the document term. All the utterances before photo is shared are concatenated, tokenized and used as the query term to retrieve the image.

\section{Experiments}

\subsection{Setup}
The maximum sequence length of BERT, ALBERT, and T5 for the photo-sharing intent prediction task is 512. We choose checkpoints that achieve the best F1 score on the dev set for evaluation on the test set.

For our dual encoder model, the maximum sequence length of BERT is 128, the dimension of the joint image-text embedding space $H$ is 512, and margin parameter $\alpha$ is 0.2 for all the experiments. All parameters are trainable. We use the Adam optimizer ($\beta1 = 0.9$, $\beta2 = 0.999$) and a learning rate
that starts at 5e-5 and decays by 0.1\% every 1000
steps. 
The models are trained on 32-core pod slices of Cloud TPU V3 Pod, with a per-replica batch size of 4. The loss is computed on item pairs
aggregated from all replicas, which is ovegr the
global batch of 128 samples in this case.

For VSE++ and SCAN models, as GRU is not a pretrained encoder, directly training them on PhotoChat yields unpleasant results. As such, we first train them on MSCOCO and finetune them on PhotoChat for 20 epochs. We utilize the same setting as the single models that are reported to perform the best on the image-retrieval task on MSCOCO; more specifically, \textit{VSE++ (ResNet, FT)} and \textit{SCAN t-i AVG ($\lambda1 = 9$)} following the annotations in the original papers. 



\subsection{Results of intent prediction}

\begin{table}[t]
\caption{Experimental results of the baseline models for the photo-sharing intent prediction task. All numbers are in percentage.} \label{tab:class_results}
\vspace{-2mm}
\centering
 \small
\begin{tabular}{l|c|c|c}
\hline
{\textbf{Model}} & \textbf{F1} $\uparrow$ & \textbf{Precision} $\uparrow$ & \textbf{Recall} $\uparrow$ \\ \hline
ALBERT-base & 52.2 & 44.8 & 62.7 \\ \hline
BERT-base & 53.2 & 56.1 & 50.6 \\ \hline
T5-base & 58.1 & \textbf{58.2} & 57.9 \\ \hline
T5-3B & \textbf{58.9} & 54.1 & \textbf{64.6} \\ \hline
\end{tabular}
 \vspace{-2mm}
\end{table}

\begin{table}[t]
    \caption{Number of negative turns and positive turns in each split of the dataset for the photo-sharing intent prediction task.}
    \vspace{-2mm}
    \centering
    \small
    \begin{tabular}{c|L{2.8cm}|L{2.8cm}} \hline
        \textbf{Split} & \textbf{Number of negatives} & \textbf{Number of positives} \\ \hline
        Train & 68,795 & 10,286 \\ \hline 
        Dev & 6,802 & 1,000 \\ \hline 
        Test & 6,748 & 1,000 \\ \hline 
    \end{tabular}
    \label{tab:sample_num}
    \vspace{-2mm}
\end{table}
Table \ref{tab:class_results} presents model performance on the test set. We observe that T5 outperforms BERT and ALBERT in all metrics.
Note that our dataset suffers from class imbalance that the negative examples outnumber the positive examples 
(roughly in a ratio of 7:1)(Table \ref{tab:sample_num})
, which we suspect causes the low precision across all the models. 

\begin{figure}[t]
\centering 
   \small
\begin{minipage}{7.7cm}\vspace{0mm}    \centering
   \begin{tabularx}{\linewidth}{c|c}
   \hline \hline
   \textbf{Example 1} & \textbf{Example 2} \\ \hline
  \raisebox{-\height}[0pt][60pt]{\includegraphics[width=0.25\textwidth]{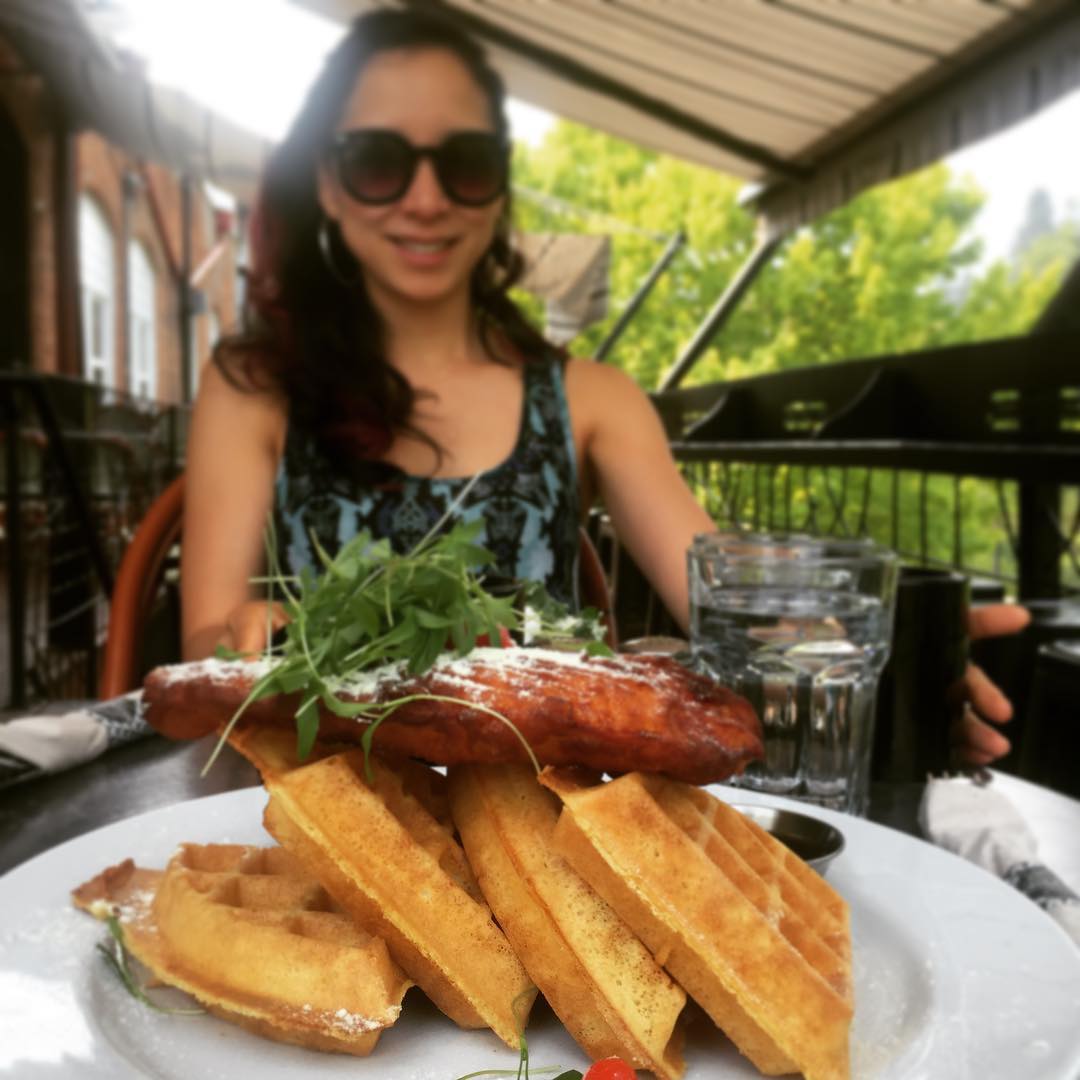}} & \raisebox{-\height}[0pt][60pt]{\includegraphics[width=0.25\textwidth]{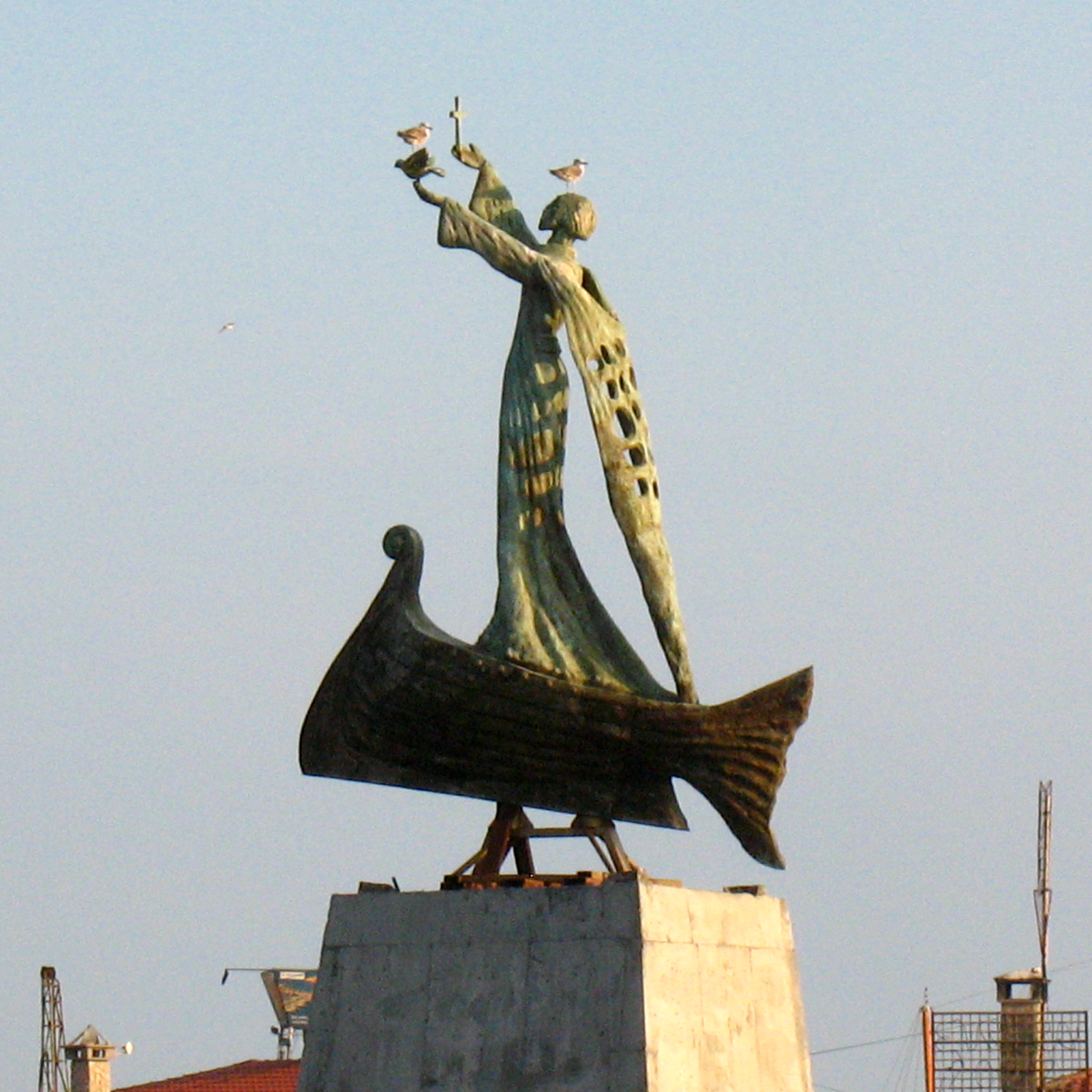}} \\
    \makecell[Xt]{
    ...\\
    \textbf{B:} That's good. I took the day off to spend with Isa. \\
    \textbf{A:} Wow \\
    \textbf{B:} It's our anniversary. \\
    \textbf{A:} Really needed sometimes \\
    \textbf{B:} We are getting brunch right now. Have you been to the blue herron cafe? \\
    \textbf{A:} no I haven't. \\
    \textbf{B:} {\textcolor{red}{\ul{\textbf{They have a beautiful balcony.}}}} \\
    \textbf{A:} tell me about it anything to share? \\
    \textbf{B:} {\textcolor{blue}{\ul{\textbf{Check out these amazing waffles!}}}} \\
    }& \makecell[Xt]{
    ... \\
    \textbf{B:} Pretty good, I spent the day at the beach with my family \\
    \textbf{A:} that sounds fun where at? \\
    \textbf{B:} Spain. They had many statues out on the beach \\
    \textbf{A:} I love the beach wow sounds beautiful!!! \\
    \textbf{B:} {\textcolor{red}{\ul{\textbf{They were really pretty}}}} \\
    \textbf{A:} did you take pics?  \\
    \textbf{B:} {\textcolor{red}{\ul{\textbf{I think so... There was this one sculpture that was unique... and the birds seemed to like it too haha}}}} \\
    \textbf{A:} {\textcolor{blue}{\ul{\textbf{oh let me see that!}}}} \\
    }\\ \hline \hline
    \end{tabularx}
\vspace{-2mm}
\caption{Predictions by T5-3B model for the intent prediction task. Turns with underline are predicted as positive. False positives are marked in red while true positives are marked in blue. Best viewed in color.}\label{fig:intent-fp-example}
\end{minipage}
\vspace{-5mm}
\end{figure}

Figure~\ref{fig:intent-fp-example} shows examples of the prediction by T5-3B model. 
Though a few turns are falsely predicted as positive (e.g. \emph{``They were really pretty."} and the second to last turn in example 2), it's possible for the speaker to share the photo after this turn in real life, indicating that when to share a photo is subjective and the model may be more viable than the low precision would suggest.
We also anticipate if the model has access to the set of photos the speaker can share, the accuracy can be elevated. In this case, the model will be able to infer that the photo in example 1 and 2 of Figure \ref{fig:intent-fp-example} are more likely to follow utterances about food and statues.

\subsection{Results of image retrieval}

\begin{table*}[t]
\caption{Experimental results of the baseline models on image retrieval task. \textit{DE} stands for our proposing dual encoders. \textit{DE$_{img}$} only uses the image pixel values and \textit{DE$_{label}$} only uses image labels to extract image features. \textit{DE*} is the model pretrained on MSCOCO. All numbers are in percentage.} \label{tab:results}
\centering
\small
\begin{tabular}{l|c|c|c|c|c}
\hline
{\textbf{Model}} & \textbf{Loss function} & \textbf{R@1} $\uparrow$ & \textbf{R@5} $\uparrow$ & \textbf{R@10} $\uparrow$ & \textbf{Sum(R@1, 5, 10)}$\uparrow$ \\ \hline
BM25 & - & 6.6 & 15.4 & 23.0 & 45.0 \\ \hline  \hline
DE$_{label}$(Bert-base) & CE & 6.7 & 22.1 & 31.2 & 60.0 \\ \hline
DE$_{img}$(ResNet-50) & CE & 6.7 & 21.9	& 32.3 & 60.9  \\ \hline
DE$_{img}$(ResNet-152) & CE & 6.8 & 24.0 & 34.3 & 65.1 \\ \hline
DE(ResNet-152, Bert-base) & CE & 8.1 & 23.7 & 34.6 & 66.4 \\ \hline
DE*(ResNet-152, Bert-base) & SH & 8.0 & 22.0 & 31.0 & 61.0 \\ \hline
DE*(ResNet-152, Bert-tiny) & SH & 7.1 & 23.3 & 33.0 & 63.4 \\ \hline
DE*(ResNet-152, Bert-base) & CE & 8.5 & 26.1 & 35.3 & 69.9 \\ \hline
\textbf{DE*(ResNet-152, Bert-tiny)} & CE & 9.0 &  \textbf{26.4} & \textbf{35.7} & \textbf{71.1} \\ \hline 
VSE++ & MH & \textbf{10.2} & 25.4 & 34.2 & 69.8  \\ \hline \hline
\textbf{SCAN} & MH & \textbf{10.4} & \textbf{27} & \textbf{37.1} & \textbf{74.5} \\ \hline
\end{tabular}
\vspace{-3mm}
\end{table*}

Table~\ref{tab:results} lists the experimental results on PhotoChat. Our dual encoder model is denoted as \textit{DE}. \textit{DE$_{img}$} and \textit{DE$_{label}$} are the ablation models that only take the image $a_i$ or image labels $l_i$ as the input compared to the default architecture in Figure~\ref{fig:model}. 
CE, SH, MH represents cross entropy loss, hinge loss, and hinge loss using hard negatives. We attempt training \textit{DE} on MSCOCO first and finetuning it on PhotoChat. These models are specially annotated with \textit{*}. We also experiment with different image encoders: ResNet-50 and ResNet-152, in combination with different label encoders: Bert-base and Bert-tiny. They are annotated in the brackets after the model names in Table~\ref{tab:results}. 
Among all the models, SCAN achieves the best performance with 10.4\% R@1, 27\% R@5, and 37.1\% R@10, which is consistent with the prior work~\citep{lee2018stacked}, demonstrating the power of the bottom-up cross attention. Among all the models that don't have cross-attention, our model \textit{DE*(ResNet-152, Bert-tiny)} performs the best and beats a strong prior work VSE++, indicating the effectiveness of using image labels in the retrieval task.

\textbf{Ablation study:} By comparing \textit{DE$_{label}$(Bert-base)} and \textit{\textit{DE$_{img}$(ResNet-152)}}, we find that using image features is more effective than using image label features, which is expected as images contain more information. Compared to the model using only image pixel values (\textit{DE$_{img}$(ResNet-152)}), adding the label features contributes to an increase of 1.3\% in sum(R@1, 5, 10) to 66.4\% (\textit{DE(ResNet-152, Bert-base)}). Pretraining the model on MSCOCO further boosts it by 3.5\% to 69.9\% (\textit{DE*(ResNet-152, Bert-base)}).

\textbf{Effect of encoders:} We observe that using a smaller model (Bert-tiny) to encode image labels yields better performance regardless of the loss function. \textit{DE*(ResNet-152, Bert-tiny)} improves sum(R@1, 5, 10) by 1.2\% compared to \textit{DE*(ResNet-152, Bert-base)} when using cross entropy loss and 2.4\% when using hinge loss. The reason might be that labels are a compact list of tokens and thus, using a smaller model alleviate the problem of overfitting. On the other hand, using a larger image encoder ResNet-152 produces better results that \textit{DE$_{img}$(ResNet-152)} beats \textit{DE$_{img}$(ResNet-50)} in sum(R@1, 5, 10) by 4.2\%.

\textbf{Effect of loss function:} Our dual encoders work significantly better with cross entropy loss than hinge loss and their gap is about 8\% in sum(R@1, 5, 10) as we compare the results of \textit{DE*(ResNet-152, Bert-base)} and \textit{DE*(ResNet-152, Bert-tiny)} models under different loss functions.

\begin{figure}[t]
  \centering
  \includegraphics[width=0.95\linewidth]{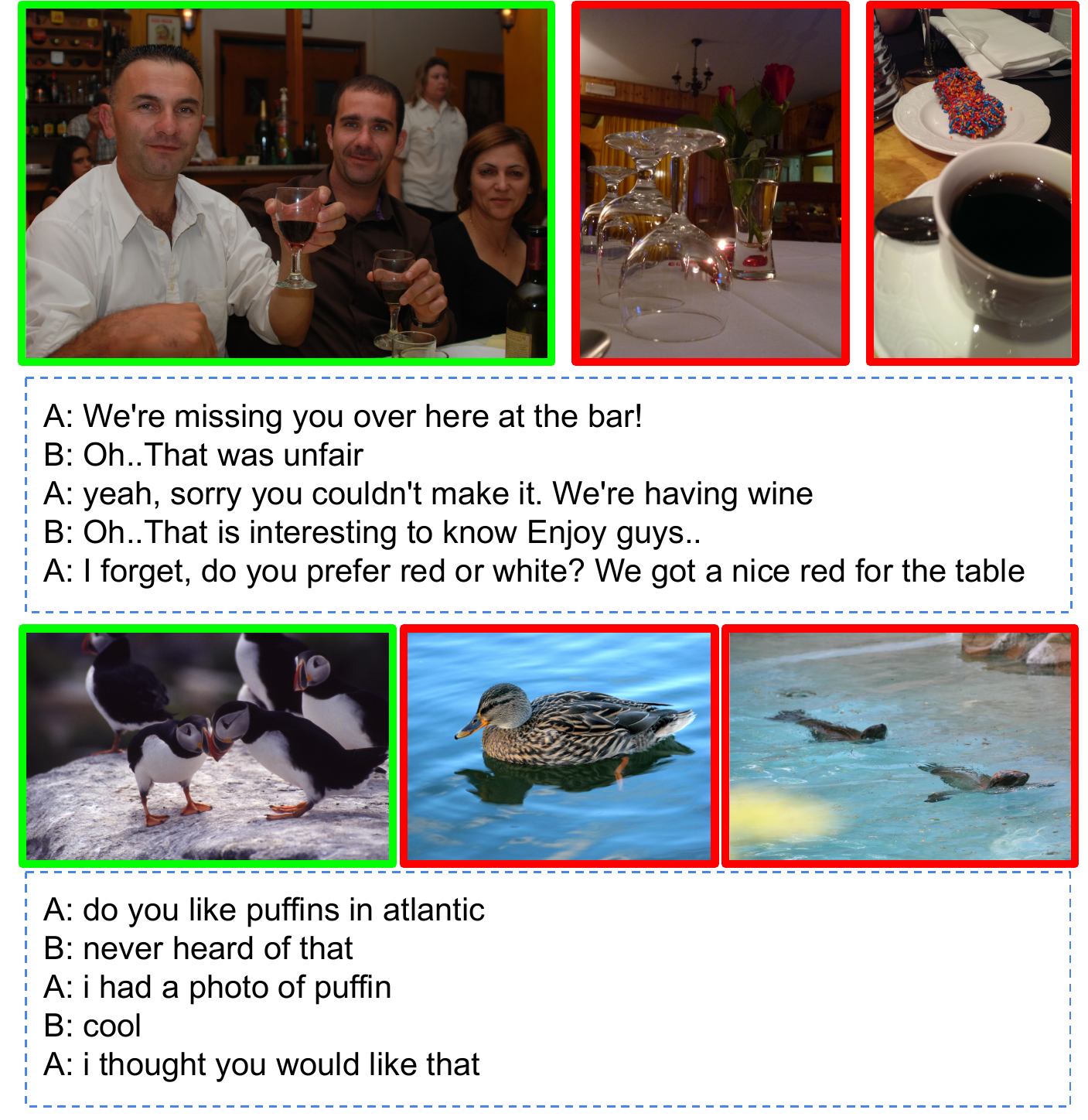}
  \caption{Predictions by \textit{DE*(ResNet-152, Bert-tiny)} for the image retrieval task. For each dialogue query, we show the groundtruth (first image in green) and the top-2 ranked images (in red). Best viewed in color.}
  \label{fig:retrieval_ex}
  \vspace{-5mm}
\end{figure}

\begin{figure*}[h!]
  \centering
  \includegraphics[width=0.99\linewidth]{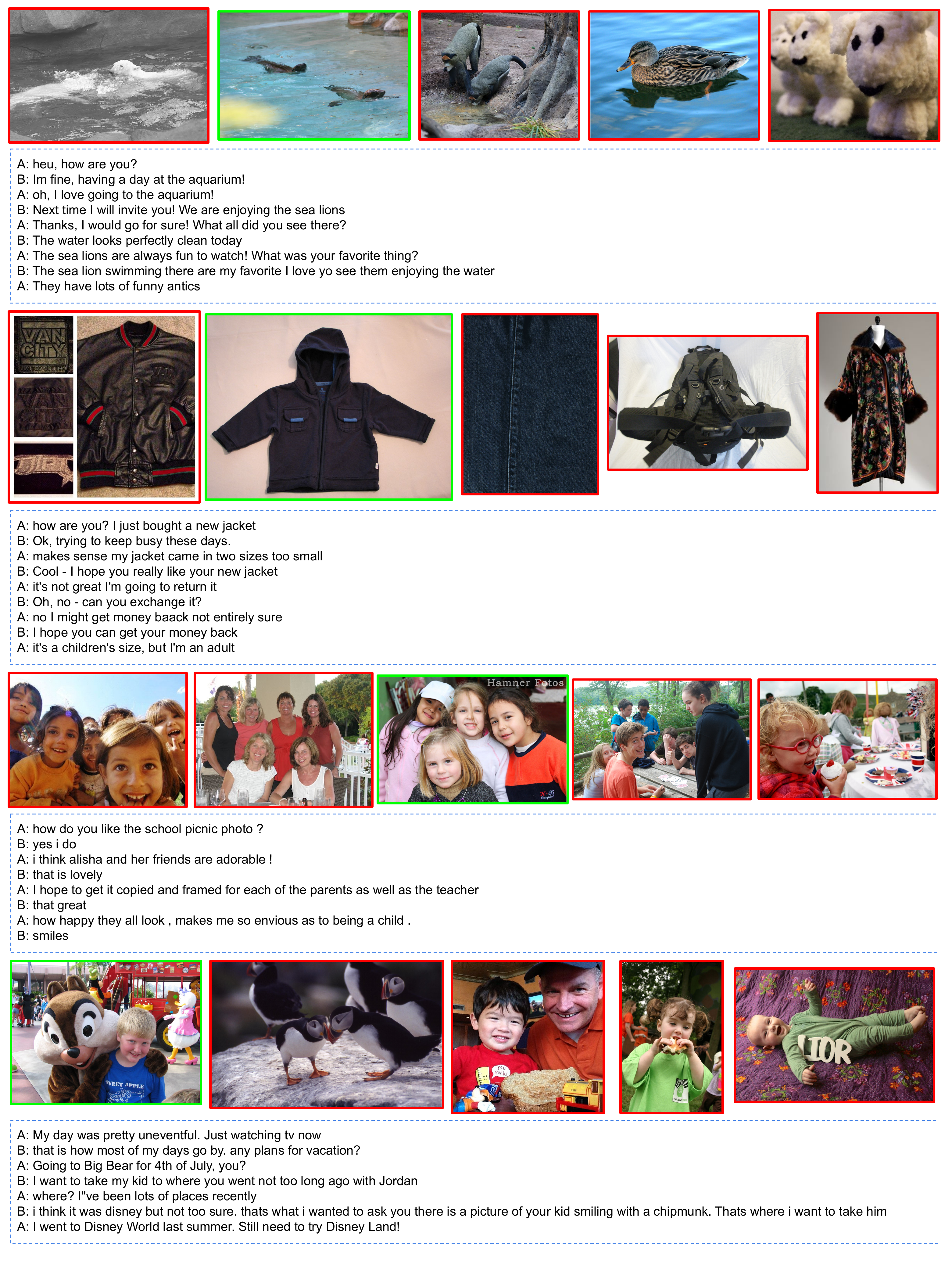}
    \vspace{-5mm}
    \caption{Predictions by \textit{DE*(ResNet-152, Bert-tiny)} for the image retrieval task. For each dialogue query, we show the top-5 ranked images from left to right. The ground-truth image is marked in green while the others are in red. Best viewed in color.}
    \label{fig:more_results}
    \vspace{-5mm}
\end{figure*}

\textbf{Error analysis:} Figure~\ref{fig:retrieval_ex} shows the qualitative results of \textit{DE*(ResNet-152,  Bert-tiny)} given a text query. In the first example, the model ranks the relevant images of wine glasses and black tea at top instead of the groundtruth image where a man is holding a wine glass, which is easy to be neglected. In the second example, the model fails to distinguish puffins with ducks and infer the background from keyword ``atlantic". It illustrates the challenge of the image retrieval task under the dialogue context that it requires a model to pay attention to the details and the event, as discussed in Section~\ref{sec:intro}. Figure \ref{fig:more_results} presents more prediction results including some wrong predictions by the model.

\section{Conclusion}
We collected a 12k high-quality dialogue dataset that contains photo sharing activity via crowd-sourcing. To facilitate research on building intelligent photo-suggest system, we have introduced two new challenging tasks that aim at improving the photo-sharing experience: photo-sharing intent prediction task and image retrieval task. That is, when given a dialogue, the system should predict whether the user has the intention to share the photo and which photo is suitable to be shared. We built baseline models for both tasks and report their performance with detailed analysis. 

Besides the proposed two new tasks, our dataset can potentially be used in other dialogue related tasks, such as dialogue generation in the multimodal dialogues, as well as inspiring new research topics, such as composing automatic reply to the photos sent from others. 
We hope our dataset and modeling work can be beneficial for studies that focus on the interplay between image and dialogue.

\small
\section*{Acknowledgments}

We thank Pranav Khaitan and Blaise Aguera y Arcas for the support and assistance; Yinfei Yang,  David Bieber for reviewing the draft and providing the feedback; Janel Thamkul and Tulsee Doshi for doing the legal review of the dataset.

\normalsize

\bibliography{anthology,acl2021}
\bibliographystyle{acl_natbib}

\appendix
\section{Dataset Creation \& Details}\label{sec:dataset_app}
\begin{figure}[h]
  \centering
  \includegraphics[width=0.90\linewidth]{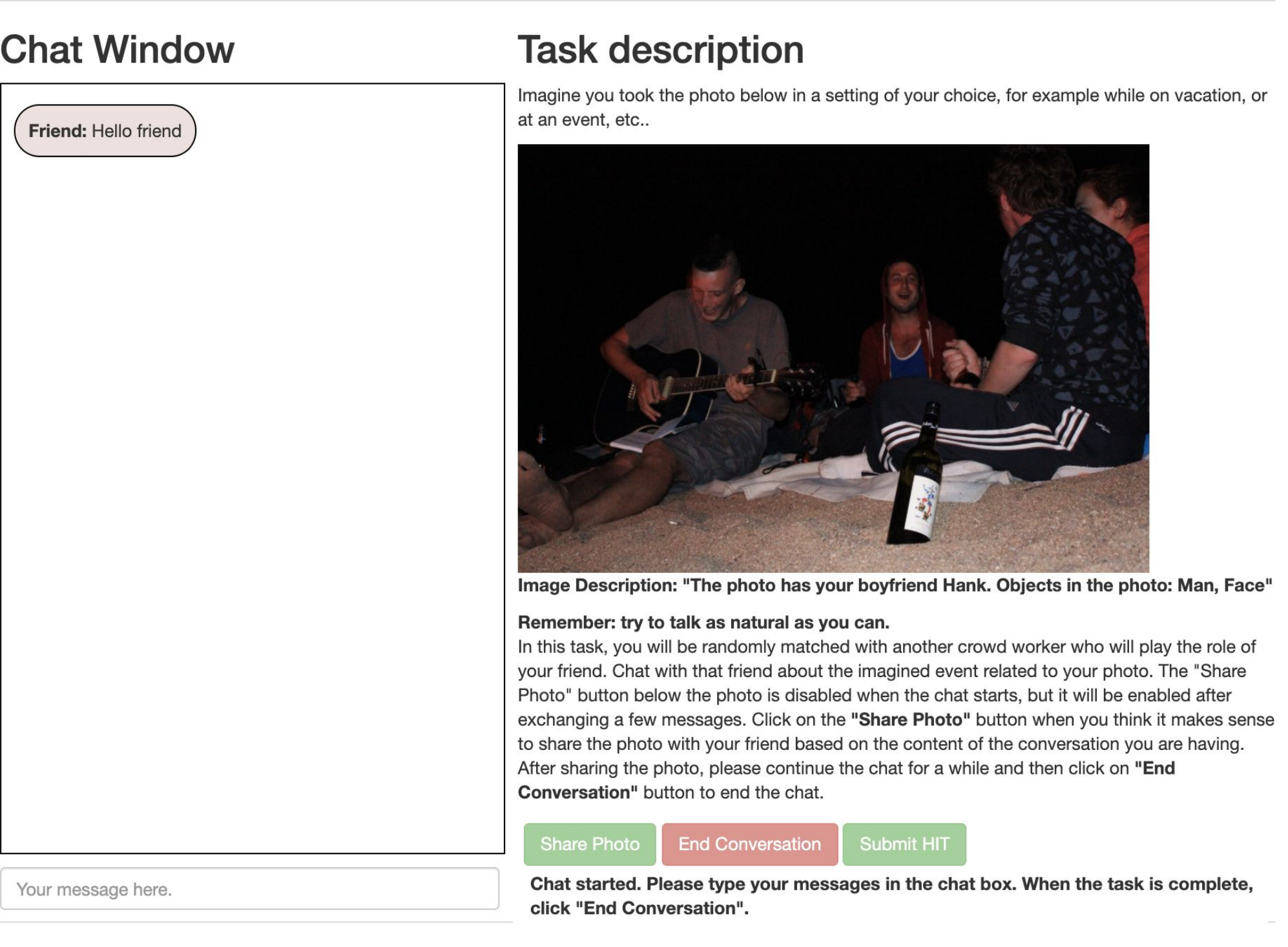}
  \caption{Website interface of the conversation generation task. It is only visible to the side who shares the photo.}
  \label{fig:skoda_1}
  \setlength{\belowcaptionskip}{-10pt}
\end{figure}

\begin{figure}[h]
  \centering
  \includegraphics[width=0.90\linewidth]{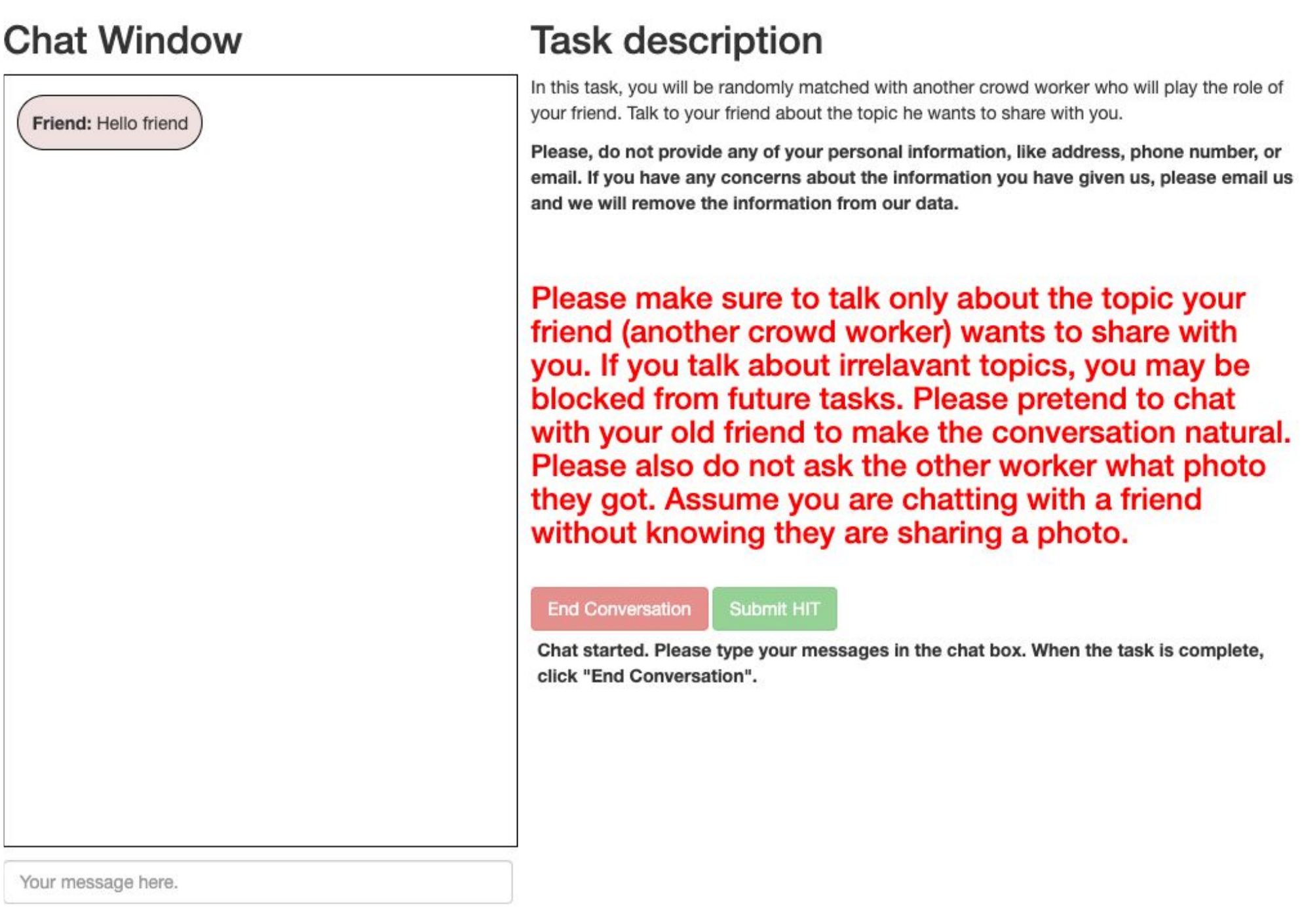}
  \caption{Website interface of the conversation generation task. It is only visible to the side who receives the photo.}
  \label{fig:skoda_2}
  \setlength{\belowcaptionskip}{-10pt}
\end{figure}

The website interfaces used to collect dialogues are presented in Figure \ref{fig:skoda_1} and \ref{fig:skoda_2}.

Table~\ref{tab:objects} shows the 89 object labels that we used to select the photos from Open Image Dataset for generating dialogues.
\begin{table}[tb]
    \centering
    \small
    \caption{Object labels we use for image filtering.}
    \begin{tabular}{C{1cm}|C{5.8cm}} \hline
        \textbf{Category} & \textbf{Object labels} \\ \hline
        People & Woman, Man, Girl, Boy, Human body, Face \\ \hline 
        Food & Bagel, Baked goods, Beer, Bread, Burrito, Cake, Candy,
        Cheese, Cocktail, Coffee, Cookie, Croissant, Dessert,
        Doughnut, Drink, Fast food, French fries, Hamburger,
        Hot dog, Ice cream, Juice, Milk, Pancake, Pasta, Pizza,
        Popcorn, Salad, Sandwich, Seafood, Snack, Taco, Tart,
        Tea, Waffle, Wine, Guacamole \\ \hline 
        Animals & Animal \\ \hline
        Products & Alarm clock, Backpack, Blender, Banjo, Bed, Belt,
        Computer keyboard, Computer mouse, Curtain, Guitar,
        Hair dryer, Hair spray, Harmonica, Humidifier, Jacket,
        Jeans, Dress, Earrings, Necklace, Fashion accessory,
        Bicycle, Blender, Calculator, Camera, Food processor, Jug,
        Mixing bowl, Nightstand, Oboe, Oven, Paper cutter,
        Pencil case, Perfume, Pillow, Personal care, Pizza cutter,
        Pressure cooker, Printer, Refridgerator, High heels,
        Skateboard, Slow cooker, Teddy bear, Teapot, Vase,
        Wall clock \\ \hline
    \end{tabular}
    \label{tab:objects}

\end{table}

\end{document}